\documentclass{aa} 
\usepackage{graphics,latexsym,amssymb,times,psfig} 
\def\gsim{ \lower .75ex \hbox{$\sim$} \llap{\raise .27ex \hbox{$>$}} } 
\def\lsim{ \ls\csiower .75ex\hbox{$\sim$} \llap{\raise .27ex \hbox{$<$}} } 
 
\begin{document} 
 
\title{Internal shocks and the blazar sequence}
\subtitle{Low and intermediate power BL Lac objects} 

\author{
Dafne Guetta \inst{1,}\inst{2},
Gabriele Ghisellini \inst{3},
Davide Lazzati \inst{4} and
Annalisa Celotti \inst{5} 
} 
 
\offprints{D. Guetta; dafne@arcetri.astro.it}
\institute{ INAF--Osservatorio Astrofisico di Arcetri, Largo E. Fermi
5, I--50125 Firenze, Italy; \and JILA, University of Colorado,
Boulder, CO 80309, USA; \and INAF--Osservatorio Astronomico di Brera,
via Bianchi 46, I--23807 Merate, Italy; \and Institute of Astronomy,
Madingley Road, CB3 0HA, Cambridge, UK; \and SISSA/ISAS, via Beirut
2-4, I--34014 Trieste, Italy.}

\titlerunning{Internal shocks in blazars}  
\authorrunning{Guetta et al.}

\abstract{ We consider internal shocks as the main dissipation
mechanism responsible for the emission in blazars and show that it can
satisfactorily account for the properties of all blazars. In
particular, we extend previous work (Spada et al. 2001) on powerful
objects, to intermediate (BL Lac) and low power sources (Mkn 421), in
order to reproduce the whole of the blazar sequence.  The model
self-consistently treats the dynamics, spectral emission and its
variability.  The key parameters driving the phenomenological sequence
are the jet power and the properties of the broad line region, which
regulate the cooling efficiency of the emitting particles and thus the
shape of the spectral energy distribution. By assuming that the
remaining parameters are similar for all objects it has been possible
to reproduce the full range of the observed spectral ``states". A more
detailed comparison of the variability properties shows (for Mkn 421)
a good agreement in the X--ray band, while in the optical the
simulated flux appears to be too variable.  For BL Lac lags ($\sim$ 10
days) are predicted between the $\gamma$--rays and the infrared
emission.
}  
\maketitle  

\section{Introduction} 
The discovery that blazars are strong $\gamma$--ray emitters together
with the results of multiwavelength campaigns have allowed to deepen
our knowledge on these objects. Their Spectral Energy Distribution
(SED) is characterized by two broad peaks (Fossati et al. 1998)
strongly variable on different timescales (Wagner \& Witzel 1995; Ulrich,
Maraschi \& Urry 1997).
 
Two main radiation processes dominate the emission, namely synchrotron
at low frequencies and -- plausibly -- inverse Compton at high
energies (see e.g. Sikora 1994 for a review).  The relative luminosity
in the two peaks and their peak frequency appear to be functions of
the total power (Fossati et al 1998, Ghisellini et al. 1998),
resulting into a sequence for the whole of the blazar population
properties, ranging from powerful, low frequency peak, through
intermediate, to low power, high frequency peak (blue) blazars (see
however Padovani et al. 2003).  The blazar emission is variable on
energy dependent timescales, typically of weeks--months in the
radio and of the order of a day in the $\gamma$--ray band.
 
Several studies, mainly based on the modeling of the SED, consistently
derived the physical parameters of the emitting region. However key
issues in the understanding of relativistic jets remain open, most
notably the jet energetics and the particle acceleration process(es).
In order to explore these issues and their relationship we have
(quantitatively) considered a scenario in which the plasma conditions
and their variability are not treated as free parameters, but follow
from the jet dynamics, thus relating the observed emission properties
with the energy transport along the jet.  Such a scenario assumes that
internal shocks are responsible for the dissipation within jets (Rees
1978 and then mostly explored for Gamma--Ray Bursts, Rees \&
M\'esz\'aros 1994).
 
The key assumption of the model is that the energy is channeled into
jets in an intermittent way by the central engine, though such a time
dependent process cannot be inferred from first principles.  Different
parts of the jet (`shells') moving at different speeds can collide
giving rise to shocks and dissipation as non--thermal radiation. The
mechanism has a limited efficiency (unless the contrast in Lorentz
factors between different shells is extremely large, see Beloborodov
2000; Guetta, Spada \& Waxman 2001, but also Ghisellini 2002), which
has to be indeed the case for blazars since most of the energy
propagates up to the extended radio lobes.
 
Beside the low efficiency, the internal shock scenario can naturally
account for other blazar properties. It predicts that jets become
radiative at $\gsim 10^{16}- 10^{17}$ cm from the central engine,
implying a minimum distance from the accretion disk and a minimum
dimension for the $\gamma$--ray source, as required by observations
(in order to avoid copious pair production; see Ghisellini \& Madau
1996).  Furthermore, successive collisions taking place at larger
distances have reduced efficiency -- since the Lorentz factor contrast
of the colliding shells decreases -- explaining why the jet luminosity
decreases with distance. Finally, the internal shock scenario appears
a promising non steady state model to account for the observed large
amplitude variability.
 
A detail study of the predictions of this model via numerical
simulations has been carried out by Spada et al. (2001; hereafter S01)
for powerful blazars (specifically 3C 279).  As the SED of this object
does not represent the whole of the blazar family mentioned above, we
focus here on two well studied sources representative of extremely
blue and intermediate blazars, namely Mkn 421 and BL Lac itself.
Rather than reproducing in detail particular spectra of specific
objects, the aim of this work is to determine whether the proposed
scenario, including the dynamics and emission properties, i) can
account for the different SED along the blazar sequence, ii) under
which hypothesis this can occur and iii) whether the internal shock
model can reproduce the observed variability behavior.
 
The outline of the paper is the following.  In \S 2 we describe the
hypothesis on the wind dynamics and on the radiative properties of the
shocked plasma.  The results -- specifically referring to the sources
Mkn421 and BL Lac -- are presented in \S 3.  In \S 4 we draw our
conclusions.

\section{Internal shocks and radiated spectra} 
 
\subsection{Physical scenario and dynamics} 
 
In this work we use the approximate model of an unsteady wind
described in detail in S01 and in the following we report only its
most relevant assumptions.
 
We consider a compact source, of typical dimension $R_0\sim 10^{14}$
cm, which produces an unstable relativistic wind characterized by an
average luminosity $L_{\rm w}$.  The emission from the wind is
obtained by adding pulses radiated in a series of internal shocks that
occur in the outflow (Daigne \& Mochkovitch 1998; Spada, Panaitescu \&
M\'esz\'aros 2000; Guetta, Spada \& Waxman 2001).
 
The wind flow is described as a sequence of $N=t_{\rm w}/t_{\rm v}$
shells, where $t_{\rm w}$ is the overall duration of the wind emission
and $t_{\rm v}\ll t_{\rm w}$ is the average interval between two
consecutive shells.  Each shell is characterized by four parameters:
 
\begin{enumerate} 
 
\item the ejection time $t_{\rm j}$, where the subscript j denotes the 
$j^{\rm th}$ shell, 
 
\item the Lorentz factor $\Gamma_{\rm j}$, 
 
\item the mass $M_{\rm j}$, 
 
\item the width $\Delta_{\rm j}$ as measured in the lab frame. 

\end{enumerate} 
 
The Lorentz factors of the shells are randomly extracted from a
uniform distribution between $\Gamma_m$ and $\Gamma_M$.  The masses
are also randomly extracted from a uniform distribution with an
average value $M_t/N$, where the total mass in the wind $M_t$ is
normalized to the jet energy budget.  The time intervals $t_{j+1}-t_j$
are drawn randomly from a uniform distribution with an average value
$t_v$.  The initial shell width $\Delta_j$ is of the same order of
$R_0$ and the initial internal energy $U_j$ is negligible (all of the
energy is in the form of bulk kinetic energy).
 
The dynamics of the wind expansion is characterized by a series of
two--shell collisions in which the faster shells catch up with the
slower ones. The model computes the collision time of two shells and
replaces them with a new shell having mass, Lorentz factor, and
internal energy given by the conservation laws for inelastic
collisions (for details see S01).  Before colliding, the shells are
assumed cold with constant thickness $\Delta=R_0$ but after collision
they get heated and thus assumed to expand at the (comoving) sound
velocity.  Adiabatic losses during the expansion are taken into
account but the bulk Lorentz factor is kept constant since the
internal energy is always much smaller than the bulk kinetic one.  The
magnetic field is generated at each collision without keeping memory
of any field generated in previous collisions.
 
In each collision a forward (FS) and a reverse (RS) shock are formed
and propagate into the front and the back shell, respectively. The
plasma parameters behind each shock are determined by the jump
conditions in mildly relativistic shocks, namely continuity in the
energy density and velocity across the contact discontinuity
separating the two shells (Panaitescu \& M\'esz\'aros 1999).  We also take
into account the compression of the merged shell width due to the
propagation of the FS and RS.
 
The energy released in each shock is assumed to be distributed among
protons, electrons and magnetic field with fractions $\epsilon_p$,
$\epsilon_{\rm e}$ and $\epsilon_{\rm B}$, respectively. The resulting
magnetic field strength allows us to evaluate the synchrotron and the
inverse Compton emission.

Electrons are assumed to be accelerated at relativistic energies with
a power law spectrum, $N(\gamma)$, above a (random) Lorentz factor
$\gamma_{\rm b}$.  $\gamma_{\rm b}$ is determined by requiring that
$\int N(\gamma) \gamma m_e c^2 d\gamma$ equals the fraction
$\epsilon_{\rm e}$ of the total available internal energy and
accounting for the fact that only a fraction, $\zeta_{\rm e}$, of
electrons is effectively accelerated.  We assume that below
$\gamma_{\rm b}$ electrons are accelerated with smaller efficiency,
corresponding to a flatter energy distribution (with respect to the
slope above $\gamma_{\rm b}$), assumed to be $\propto \gamma^{-1}$.
 
In conclusion, by considering the wind dynamics and shock
hydrodynamics, the energy density in non--thermal electrons and in
magnetic field and the typical random Lorentz factor $\gamma_{\rm b}$
can be determined for each collision.
 
\subsection{Local spectra} 
 
We followed the same basic assumptions of S01, which we briefly
recall:
 
\begin{itemize} 
 
\item 
The emitting zone is homogeneous with comoving volume $V=\pi \psi^2
R^2\Delta R^\prime$, where $\psi$ is the half--opening angle of a
conical jet and $\Delta R^\prime$ is the comoving thickness of the
shell at the collision time, whose expansion while emitting is
neglected for simplicity.

\item 
The relativistic particles, embedded in a tangled magnetic field, have
the same energy distribution throughout the region corresponding to
the shell--shell interaction. This simplification is justified as we
are interested in the ``average" spectrum, but does not account for
spectral details occurring on a timescale shorter than the light
crossing time.
 
\item 
Even in the case of low power BL Lacs, we consider the presence of
soft photons external to the jet (identified with the emission from
the broad line region, BLR).  The external photon
luminosity $L_{\rm ext}=a L_{\rm disk}$ is produced within $R_{\rm
BLR}$ and abruptly vanishes beyond, and corresponds to a comoving
radiation energy density $U_{\rm ext} = (17/12) aL_{\rm disk} \Gamma^2
/(4 \pi R_{\rm BLR}^2 c)$ (e.g. Ghisellini \& Madau 1996).
 
High and low power blazars have different line/photoionizing disk
luminosities. Following Kaspi et al. (2000), $R_{\rm BLR}$ and $L_{\rm
disk}$ are related by $R_{\rm BLR}\propto L_{\rm disk}^b$ with $b\sim
0.7$.  Thus blazars with weaker broad emission lines should have
smaller BLR, implying that the first collisions occur preferentially
outside $R_{\rm BLR}$.
 
\item 
Relativistic electrons are injected with a broken power--law energy
distribution, $\propto \gamma^{-1}$ below $\gamma_{\rm b}$ and
$\propto \gamma^{-s}$ between $\gamma_{\rm b}$ and $\gamma_{\rm max}$,
with a power $L_{\rm e}$.  The equilibrium particle distribution
$N(\gamma)$, resulting from the continuous injection and cooling
processes, is determined through the following procedure.
 
\noindent 
Above the (comoving) random Lorentz factor $\gamma_{\rm cool}$,
corresponding to a cooling time comparable with the shock crossing
time $t_{\rm cross}$, the distribution is assumed to have a power law
index $p=s+1$.  Below $\gamma_{\rm cool}$ there are two different
behaviors:
 
i) if $\gamma_{\rm cool} > \gamma_{\rm b}$, we assume $N(\gamma) \propto 
\gamma^{\rm -s}$ between $\gamma_{\rm b}$ and $\gamma_{\rm cool}$, and 
$N(\gamma)\propto \gamma^{-1}$ for $\gamma < \gamma_{\rm b}$; 
 
ii) if $\gamma_{\rm cool} < \gamma_{\rm b}$, $N(\gamma) \propto 
\gamma^{-2}$ between $\gamma_{\rm cool}$ and $\gamma_{\rm b}$, and 
$N(\gamma)\propto \gamma^{-1}$ for $\gamma <\gamma_{\rm cool}$. 
 
\item 
The normalization of $N(\gamma)$ is determined according to whether
particles with $\gamma\sim \gamma_{\rm b}$ can or cannot cool in the
timescale $t_{\rm cross}$ (fast and slow cooling regime,
respectively):
 
{\bf Fast cooling regime ---} If most of $L_{\rm e}$ is radiated in a
timescale $<t_{\rm cross}$ (i.e. $\gamma_{\rm b} > \gamma_{\rm cool}$)
we apply a luminosity balance condition:
\begin{equation} 
L_{\rm e} \, =\, V m_{\rm e}c^2 \int \dot\gamma N(\gamma) d \gamma, 
\end{equation} 
where $\dot \gamma$ is the radiative cooling rate, including
synchrotron, synchrotron self--Compton (SSC) and external Compton (EC)
losses.  In order to allow for multiple Compton scatterings we a
priori calculate how many scattering orders $n_{\rm IC}$ can take
place before the Klein--Nishina regime is reached.  By considering
$\gamma_{\rm b}$ as the relevant energy
\begin{equation} 
n_{\rm IC}\, =\, {\ln (\gamma_{\rm b}/x_{\rm B}) \over \ln (4\gamma_{\rm 
b}^2/3)} \, -1, 
\end{equation} 
where $x_{\rm B}\equiv h\nu_{\rm B}/(m_{\rm e}c^2)$ and $\nu_{\rm
B}=eB/(2\pi m_{\rm e} c)$ is the (non--relativistic) Larmor frequency.
We define $U_{\rm e} \equiv L_{\rm e}/(\pi R^2 c)$ and introduce the
Comptonization parameter $y$:
\begin{equation} 
y\, \equiv \, {3\over 4} \sigma_{\rm T} \Delta R^\prime 
\int \gamma^2 N(\gamma) d \gamma. 
\end{equation} 

The equality between the injected and the radiated power in the fast
cooling regime translates into an equality between the electron and
the radiation energy densities, where the latter comprises the
synchrotron and the $n_{\rm IC}$ inverse Compton orders energy
density, i.e.

\begin{equation} 
U_{\rm e}\, = y U_{\rm B} \left( 1+{U_{\rm ext} \over U_{\rm B}} + y + 
y^2 + \, ....\, y^{n_{\rm IC}}\right). 
\end{equation} 
This provides the normalization of $N(\gamma)$, assuming the above
slopes.
 
{\bf Slow cooling regime ---} If most of the power is not radiated in
$t_{\rm cross}$, the electrons retain their energy and
\begin{equation} 
E_{\rm e} \, =\, L_{\rm e} t_{\rm cross}\, = \, m_{\rm e}c^2 V \int 
N(\gamma)\gamma d\gamma. 
\end{equation} 
Assuming the same distribution for $N(\gamma)$, its normalization can
be determined.

\item 
The slopes of the $N(\gamma)$ distribution are fixed, but the cooling
energy is found by iteration as $\gamma_{\rm cool}$ depends on the
amount of SSC radiation, which in turn depends on the exact shape of
$N(\gamma)$ (because of Klein--Nishina effects). Since $\gamma_{\rm
cool}$ determines the normalization of $N(\gamma)$, we re--iterate the
procedure described above until it converges (usually 3--4 iterations
are enough).
 
\item 
The synchrotron--self--absorption process, the inverse Compton
emission and the beaming of the radiation are treated as in S01.
 
\end{itemize} 
 
\subsection{Observed spectrum} 
 
Both the particle and radiation local spectra thus determined are not
steady, but roughly correspond to those attained after a time $t_{\rm
cross}$ after the shell--shell interaction, at the maximum bolometric
radiative output.  This spectrum -- adopted for the entire duration of
the emission -- has a normalization modulated as a function of time.
 
Each photon pulse -- with its corresponding spectrum -- starts at the
time of a collision and lasts for a duration set by the combination of
the radiative, shock crossing and angular spread timescales.  The
shape of the pulses (raise and decay in the light curves) is assumed
symmetric and linear with time if its duration is determined by
geometrical effects and/or by the shock crossing time, as generally
occurs for collisions inside the BLR (as typical of powerful blazar
like 3C279).  If the pulse duration is instead set by cooling, the
pulse still rises linearly, but decays exponentially [$\propto
\exp(-t/3 t_{\rm cool}(\gamma_{\rm b}) )$], implying that the rise and
decay times can be different.  This usually occurs in weaker blazars,
like Mkn 421.  The pulses of intermediate BL Lac objects can be
characterized by either of these pulse shapes, depending on the broad
lines intensity and thus on whether collisions take place within
$R_{\rm BLR}$.

The observed spectrum is determined by all of the photons reaching the
observer simultaneously: since these are produced at different
distances, their different light propagations times should be taken
into account.  In other words, the resulting spectrum is obtained as
the convolution of the emission produced by shocks simultaneously
active in the observer frame, taking into account the radius at which
collisions occur and the photon travel path.
 
\begin{figure} 
\psfig{figure=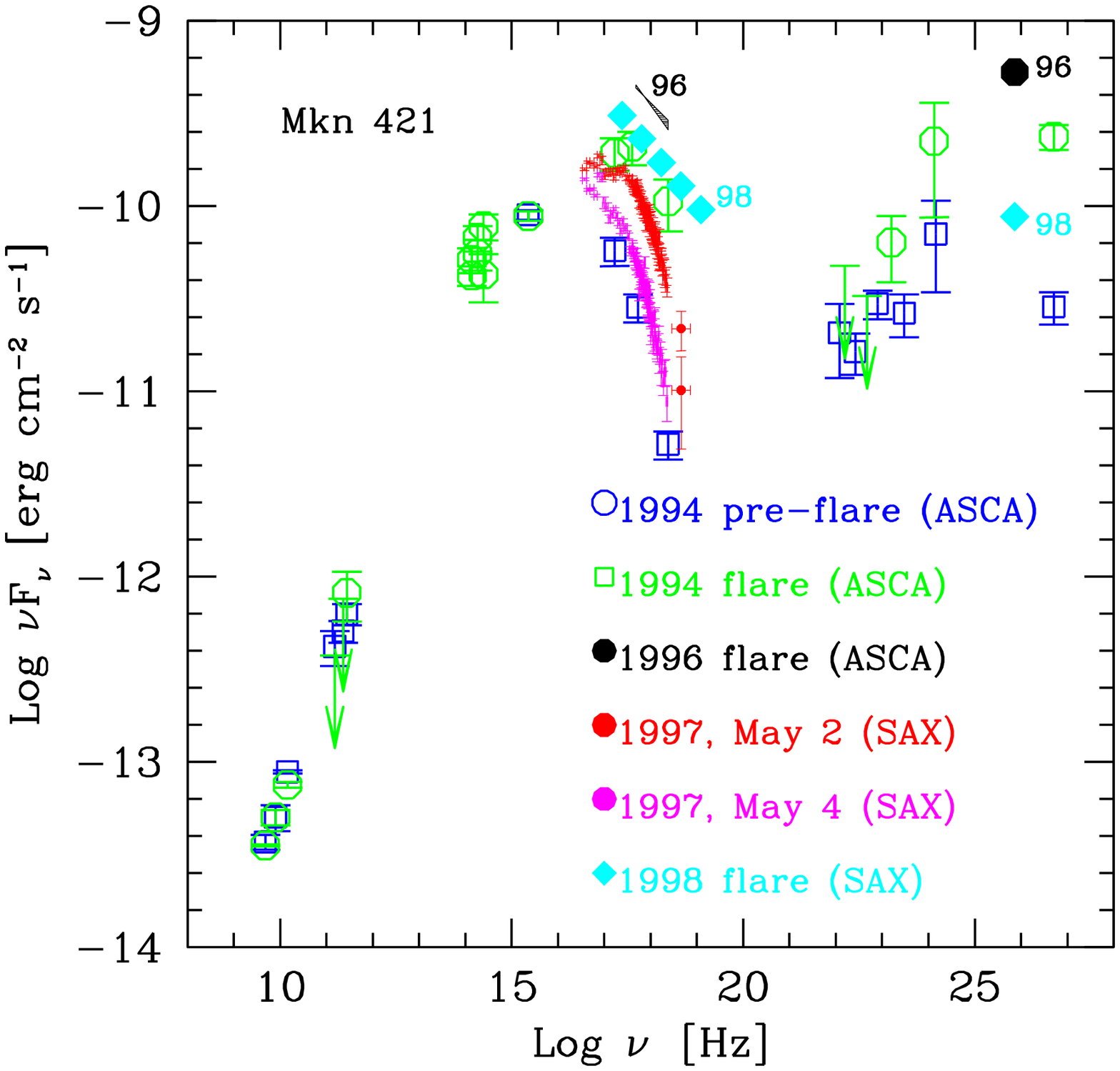,angle=0,width=9cm} 
\vskip -0.5 true cm 
\caption{The SED of Mkn421 during different observational campaigns,
as labeled. For the data points see Costamante \& Ghisellini (2002)
and references therein.}

\psfig{figure=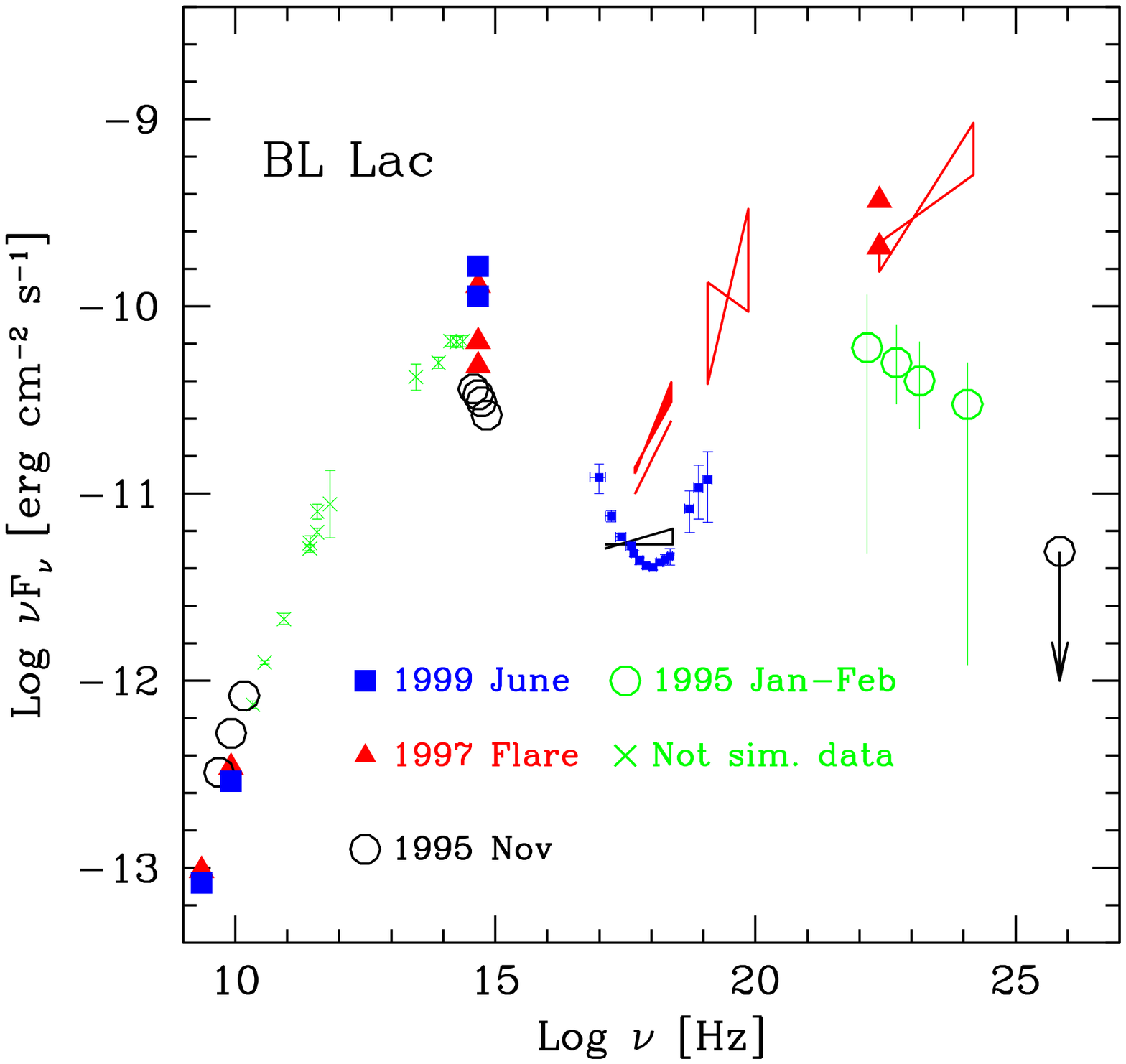,angle=0,width=9cm}
\vskip -0.5 true cm 
\caption{The SED of BL Lac during different observational campaigns,
as labeled. For the data points see Ravasio et al. (2002) and
references therein.}
\end{figure} 

\section{Results} 
 
The main aim of this work is to explore whether the internal shock
scenario can account for the difference of the SED along the blazar
sequence.  We thus a) concentrate on the global properties, rather
than to explain details of the spectrum of a specific object; b) apply
the model to Mkn 421 and BL Lac itself which, besides being among the
best studied BL Lac objects, are representative of extremely blue and
intermediate blazars, respectively, complementing the study already
performed for the powerful blazar 3C 279 (S01).
 
{\it Mkn 421 --- } This is one example of bright, extremely blue and
strong TeV emitting blazar.  The peaks of its synchrotron and inverse
Compton spectra are at $\sim$ keV and a few hundreds GeV energies,
respectively.  The bolometric (observed) power is of order $8\times
10^{45}$ erg s$^{-1}$.  The SED of coordinated simultaneous
observational campaigns of Mkn 421 are reported in Fig.~1 (Costamante
\& Ghisellini 2002).

{\it BL Lac --- } The prototype of the BL Lacertae class is an
intermediate blazars (e.g. Fossati et al. 1998).  It is a 1 Jy source,
with broad band peaks in the near IR and MeV--GeV band and its total
bolometric power is of order $\sim 10^{46}$ erg s$^{-1}$.  Despite of
BL Lac being the prototype of the BL Lac class, it sporadically showed
optical emission lines (with maximum observed EW $\sim 6$ \AA;
Vermeulen et al. 1995; Corbett et al. 2000).  In July 1997 it
underwent a major outburst, followed in the optical and by the CGRO
(EGRET), {\it Rossi}XTE and ASCA satellites.  During this flare the
entire SED dramatically changed, as illustrated in Fig. 2, with EGRET
detecting a flux ($>$100 MeV) a factor 3.5 times higher than that
observed in 1995 (Bloom et al. 1997).

\begin{figure} 
\vskip -0.5 true cm
\psfig{figure=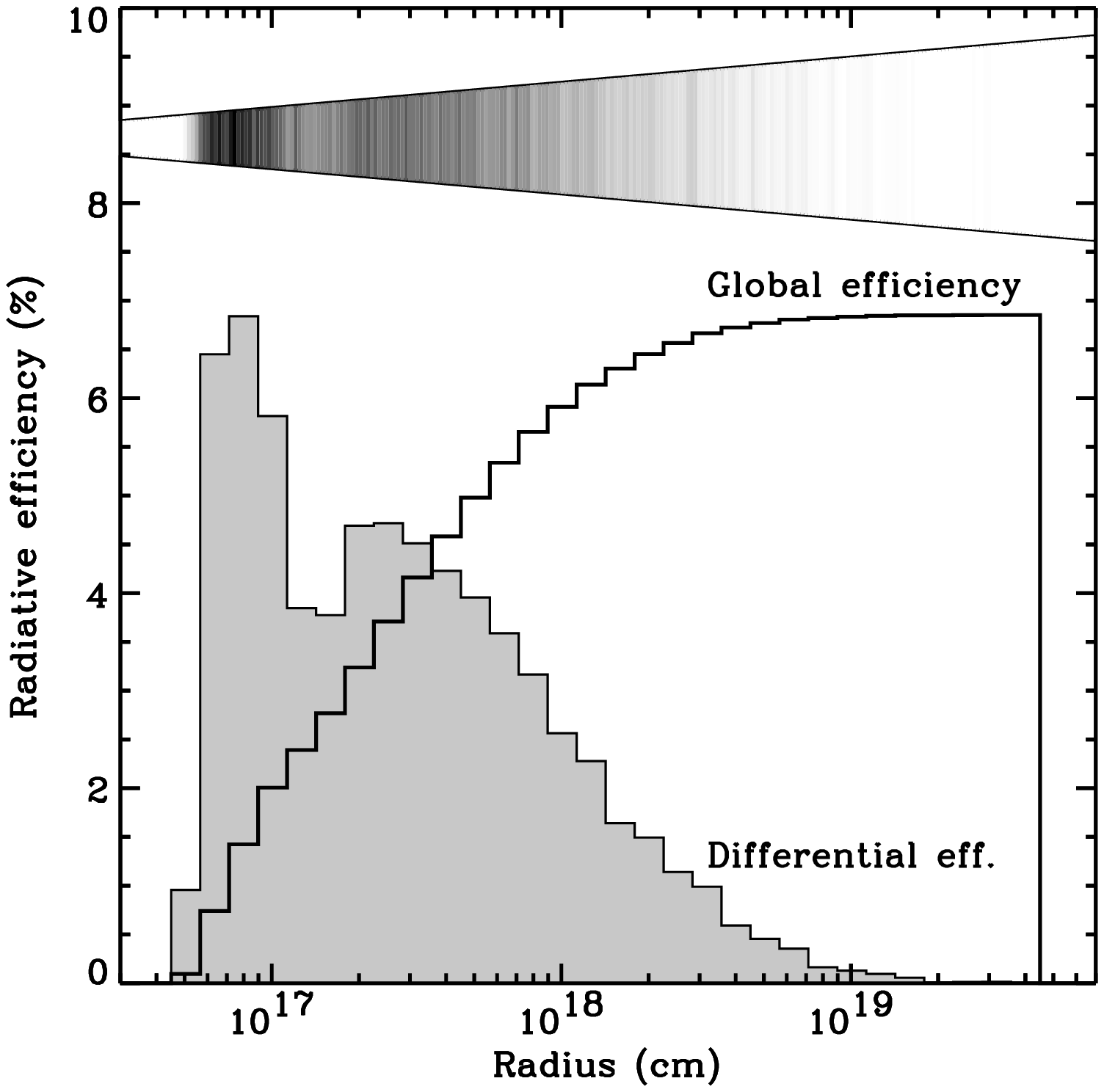,width=9cm} 
\vskip -0.5 true cm 
\caption{Radiative efficiency as a function of the collision radius
for Mkn 421.  The solid line refers to the global efficiency, i.e.
the fraction of the total wind kinetic energy ($E_{\rm w}=L_{\rm
w}\times t_{\rm w}$) radiated on scales smaller than any given radius;
the shaded histogram shows the differential efficiency, namely the
fraction of $E_{\rm w}$ radiated for a given radius interval.  The
cone--like insert in the upper part of the figure shows a grey--tone
representation of the differential efficiency of the jet: the darker
the color the higher the efficiency.}
%
%
\psfig{figure=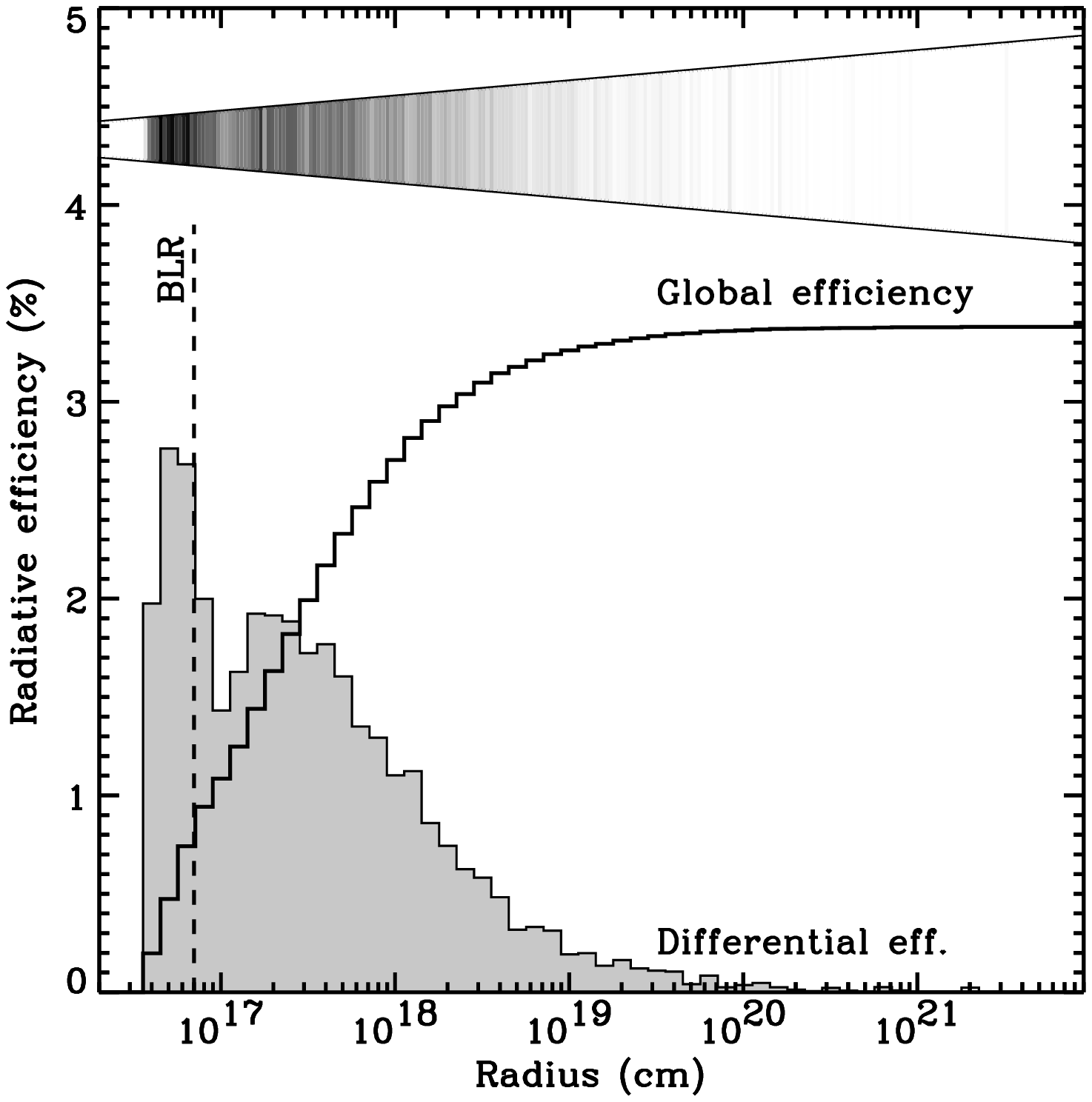,width=9cm} 
\vskip -0.5 true cm 
\caption{Radiative efficiency versus the collision radius for BL Lac.
The solid line refers to the global efficiency, i.e.  the fraction of
the total wind kinetic energy ($E_{\rm w}=L_{\rm w}\times t_{\rm w}$)
radiated on scales smaller than a given radius; the shaded histogram
shows instead the differential efficiency, namely the fraction of
$E_{\rm w}$ radiated for a given radius interval.  The cone--like
insert in the upper part of the figure shows a grey--tone
representation of the differential efficiency of the jet: the darker
the color the higher the efficiency. The vertical line indicates
$R_{\rm BLR}$.}
\end{figure} 
 
In order to reproduce the SED and variability of these two sources, we
took as an initial set of physical parameters that determined for 3C
279 (see S01), and changed as few as possible of them to account for
the different broad band SED and 'spectral states' of these sources.

It turned out that the global properties of the blazar ``sequence" can
be reproduced by only changing two key quantities, namely the jet
luminosity $L_{\rm jet}$ and the luminosity $L_{\rm BLR}$ and
extension $R_{\rm BLR}$ of the external photon field.  The other
parameters [$\epsilon_{\rm B}$, $\zeta _{\rm e}$, the slope $n$ of the
(high energy) electron distribution, the observing angle $\theta$] are
only constrained to fit the specific SED.  The remaining quantities
are at most slightly varied just to optimize the fit.  In particular
$\epsilon_{\rm e}$ has been fixed for all sources to 0.5.
In Table 1 we list all of the input parameters adopted to reproduce
the time dependent behavior of the simulated sources (for convenience
of the reader we also report the same parameters for 3C 279 from S01).
 
Let us now examine the physical quantities inferred from the modelling
and their dependence on $R$.
 
\vskip 0.3 true cm
\noindent 
{\it Efficiency ---} The radiative efficiency represents the fraction
of the dissipated energy at each location channeled into relativistic
electrons.  In absence of non--radiative losses this energy should be
all radiated.  As shown in Fig. 3 and 4 for Mkn 421 and BL Lac,
respectively, the efficiency decreases simply due to the decrease in
the contrast of bulk Lorentz factors of colliding shells.  The
apparent different efficiencies of BL Lac and Mkn 421 reflect only the
different viewing angle (chosen to reproduce the detailed properties
of the two sources).

\vskip 0.3 true cm
\noindent 
{\it Parameters profiles: $\Gamma$, $B$, $\gamma_{\rm b}$ ---} Fig. 5
and Fig. 6 (top three panels) show how the bulk Lorentz factor, the
magnetic field, the Lorentz factor $\gamma_{\rm b}$ of the injected
electrons evolve with distance from the nucleus, for Mkn 421 and BL
Lac, respectively.  Different symbols correspond to shells having
experienced a different number of collisions.  The bulk Lorentz
factors tend toward average values as result of successive collisions
and merging of the shells.  The range and the asymptotic bulk Lorentz
factor $\Gamma$ are fully consistent with the typical values inferred
from observations.  The magnetic field decreases roughly as $R^{-3/2}$
for both sources: it is worth remarking that we assume that a constant
fraction ($\epsilon_{\rm B}$) of the available energy is converted
into magnetic field and no field component survives between successive
collisions.  The minimum energy of the injected electrons, $\gamma_b
m_e c^2$, is also a decreasing function of $R$, due to the decrease in
the bulk Lorentz factors contrast, implying decreasing efficiency and
mean energy per particle.
 
In general the only difference between Mkn 421 and BL Lac is the
smaller spread of the parameters for the former, due to the fact that
in Mkn421 all the collisions occur in the same region.

\vskip 0.3 true cm
\noindent 
{\it Parameters profiles: $\gamma_{\rm peak}$, $\nu_{\rm peak}$ ---}
The two bottom panels of Figs. 5, 6 show the spatial profile of the
energy of the electrons emitting at the peak of the SED, $m_{\rm e}
c^2 \gamma_{\rm peak}$, and the corresponding observed synchrotron
peak frequency, $\nu_{\rm peak}$.
 
A significantly different behavior characterizes the two sources. In
Mkn 421 $\gamma_{\rm peak}$ is a continuous decreasing function of $R$
and attains large values.  This follows from the fact that in low
power blazars only high energy electrons can radiatively cool in a
finite injection time, thus determining high values of $\gamma_{\rm
peak}$.  Specifically, its value is determined by $\gamma_{\rm max}$
at each collision radius, as set by the spectral fit.  The behavior of
$\nu_{\rm peak}$ simply follows that of $\gamma_{\rm peak}$ and $B$.
On the contrary for BL Lac $\gamma_{\rm peak}$ (and consequently
$\nu_{\rm peak}$) shows a discontinuity.  This occurs at $R_{\rm
BLR}$: inside the BLR electrons are in the fast cooling regime and
thus $\gamma_{\rm peak}=\gamma_b$ while outside the BLR the behavior
is similar to that of Mkn 421, namely the radiative cooling time
exceeds the shock crossing time, implying $\gamma_{\rm
peak}=\gamma_{\rm max}$.  Note that for 3C 279 the BLR is even larger
and $\gamma_{\rm peak}=\gamma_b$ for almost all of the collisions.

\begin{figure} 
\psfig{figure=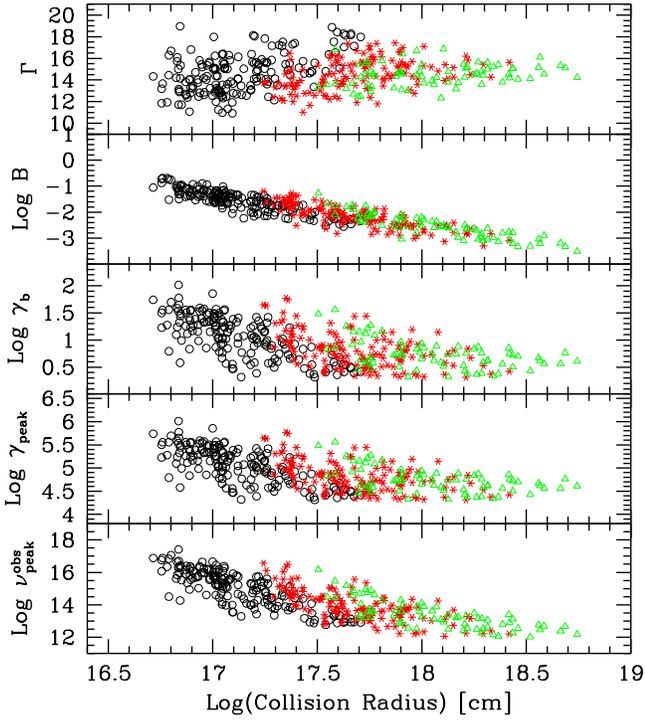,angle=0,width=10.3cm} 
\caption{ Evolution of the post--shock shell parameters as a function
of the collision radius for Mkn 421.  The upper panel shows the bulk
Lorentz factor of the merged shell after each collision; the second
panel represents the value of the generated magnetic field
($\epsilon_{\rm B} =6\times 10^{-4}$).  The third panel shows the
minimum random Lorentz factor of the injected electrons $\gamma_{\rm
b}$ (for $\epsilon_{\rm e} = 0.5$; $\zeta_{\rm e} = 0.08$).  In the
fourth and last panels $\gamma_{\rm peak}$ and $\nu^{\rm obs}_{\rm
peak}$ are plotted.  Only one tenth of the points is displayed for
clarity.  Circles refer to collisions in which none of the shells has
ever collided before, stars to collisions in which one of the shells
has collided, while triangles indicate collisions between shells that
have both collided before.  }
\end{figure} 
  
\begin{figure} 
\psfig{figure=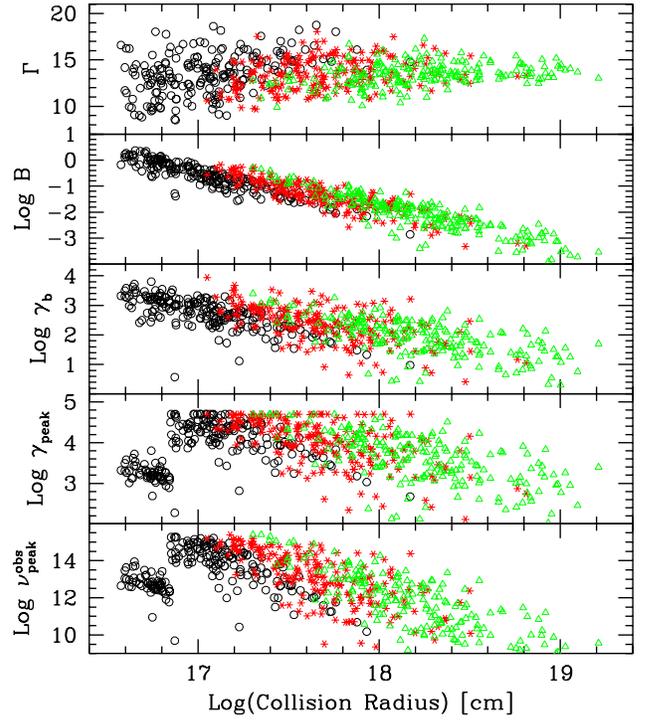,angle=0,width=10.3cm} 
\caption{ Evolution of the post--shock shell parameters as a function
of the collision radius for BL Lac.  The upper panel shows the bulk
Lorentz factor of the merged shell after each collision; the second
panel represents the value of the generated magnetic field
($\epsilon_{\rm B} =10^{-2}$).  The third panel shows the minimum
random Lorentz factor of the injected electrons $\gamma_{\rm b}$ (for
$\epsilon_{\rm e} = 0.5$; $\zeta_{\rm e} = 10^{-2}$).  In the fourth
and last panels $\gamma_{\rm peak}$ and $\nu^{\rm obs}_{\rm peak}$ are
plotted.  Only one tenth of the points is displayed for clarity.
Circles refer to collisions in which none of the shells has ever
collided before, stars to collisions in which one of the shells has
collided, while triangles indicate collisions between shells that have
both collided before.  }
\end{figure} 

\vskip 0.3 true cm
\noindent  
{\it Spectra at various $R$ ---} In order to understand how and where
the overall spectrum is generated, the spectra produced at different
radii are shown in Fig. 7 and Fig. 8: in the top panel the average
``instantaneous" spectrum of a single shell at $R$ is shown while in
the bottom panel the spectra are integrated for all of the shells
emitting at a certain time (the flux is thus weighted for the duration
of the emission in each shell at a given frequency).  While the single
shell emission is qualitatively similar for the two sources, the
integrated spectra of BL Lac do not simply reproduce the single shell
behavior.  This is due to the different (shorter) duration of each
single shell emission, due to the fast cooling of the electrons
(caused by the external Compton).  The end result is that in Mkn 421
most of the emission is produced in the inner jet, while in BL Lac the
inner part only dominates at the highest frequencies of the two
spectral peaks,
with interesting consequences on the correlations between variability
in different bands.
 
It is also worth commenting on the extreme differences in the SED of
BL Lac at different epochs (as shown in Fig. 2).  This has been
already interpreted as due to variable broad emission lines in the
spectrum (mainly $H_\alpha$ and $H_\beta$), which would cause a
variable contribution of seed photons for the external Compton
process.  The maximum observed total luminosity from the BLR is
5--8$\times 10^{42}$ erg s$^{-1}$ (see e.g. Vermulen et al., for
$H_0=65$ km s$^{-1}$ Mpc$^{-1}$).  If one adopts the correlation
(Kaspi et al. 2000) between $R_{\rm BLR}$ and the ionizing continuum
luminosity (here assumed to be $\sim$ 10 times $L_{\rm BLR}$), $R_{\rm
BLR}\sim$ 5--7 $\times 10^{16}$ cm.  This inferred distance is 
indeed similar to 
the typical distance of the first shell--shell collisions, 
suggesting that the collisions sometimes occurs within $R_{\rm BLR}$,
sometimes outside it (see also Ravasio et al. 2002, 2003).  Dramatic
spectral changes, such as those between the 1995 and the 1997 SED (see
Fig. 2) can be therefore accounted for in this scenario.

\begin{figure} 
\vskip 0.5 true cm \psfig{figure=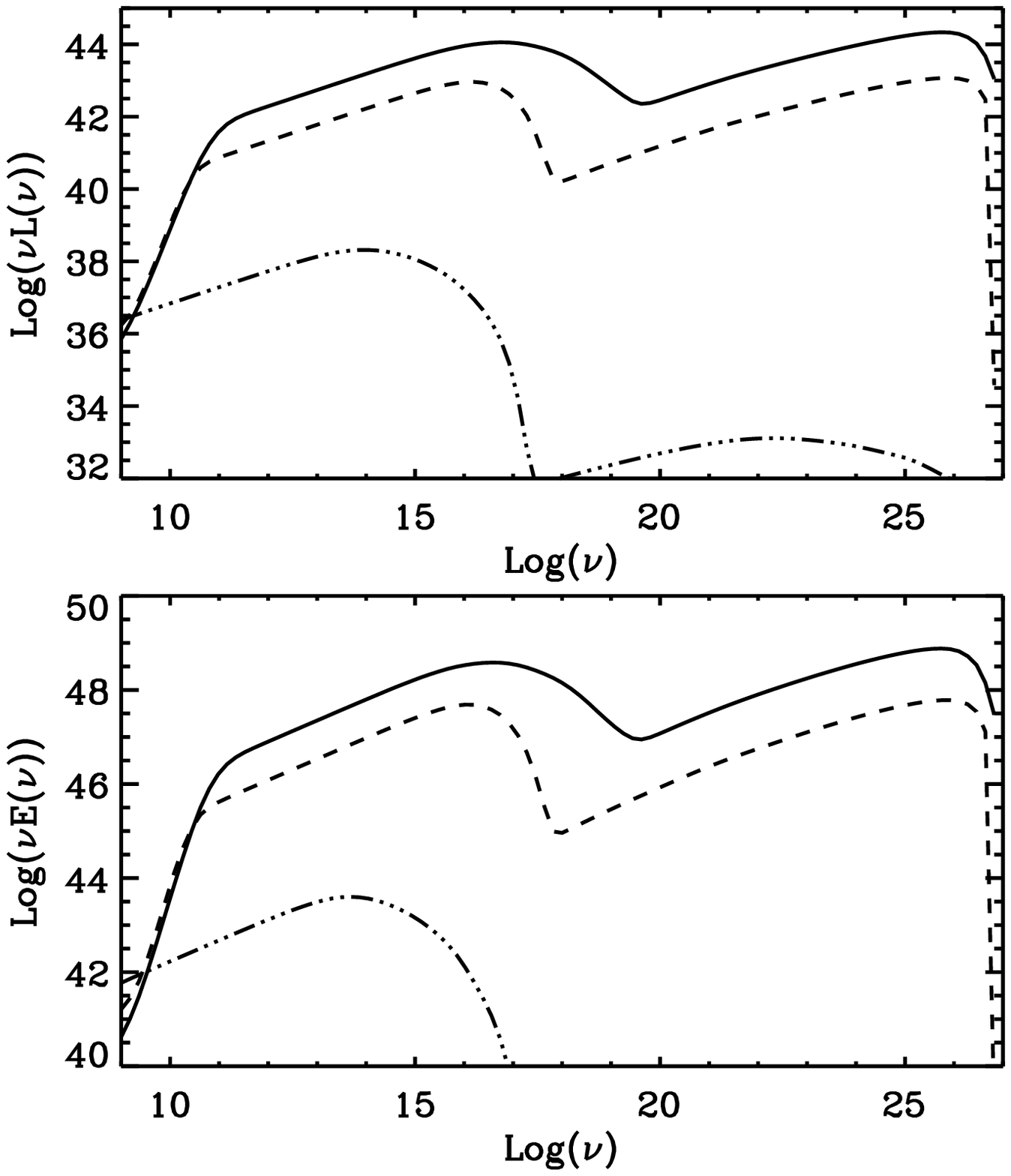,width=8cm}
\caption{ Average spectra of collisions for Mkn 421.  The upper panel
shows the average spectrum of the shells without taking into account
the duration of the emission.
In the lower panel, instead, the duration of the emission of each
shell has been taken into account.}
%
\vskip 0.5 true cm \psfig{figure=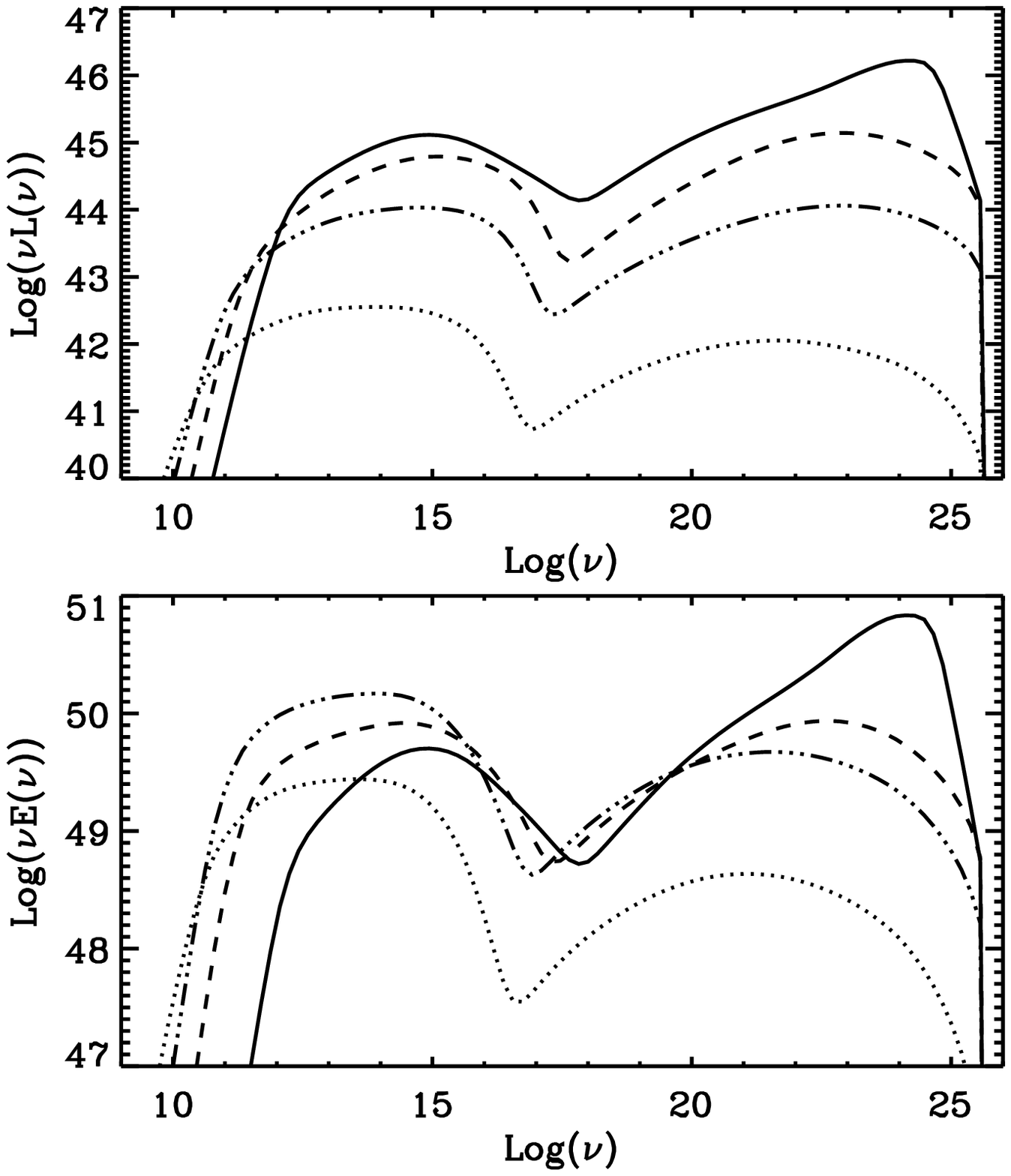,width=8cm}
\caption{ Average spectra of collisions for BL Lac.  The upper panel
shows the average spectrum of the shells without taking into account
the duration of the emission. 
In the lower panel, instead, the duration of the emission of each
shell has been taken into account.}
\end{figure} 
 
\begin{figure}
\vskip 0.5 true cm 
{\hskip 1.3 truecm 
\psfig{figure=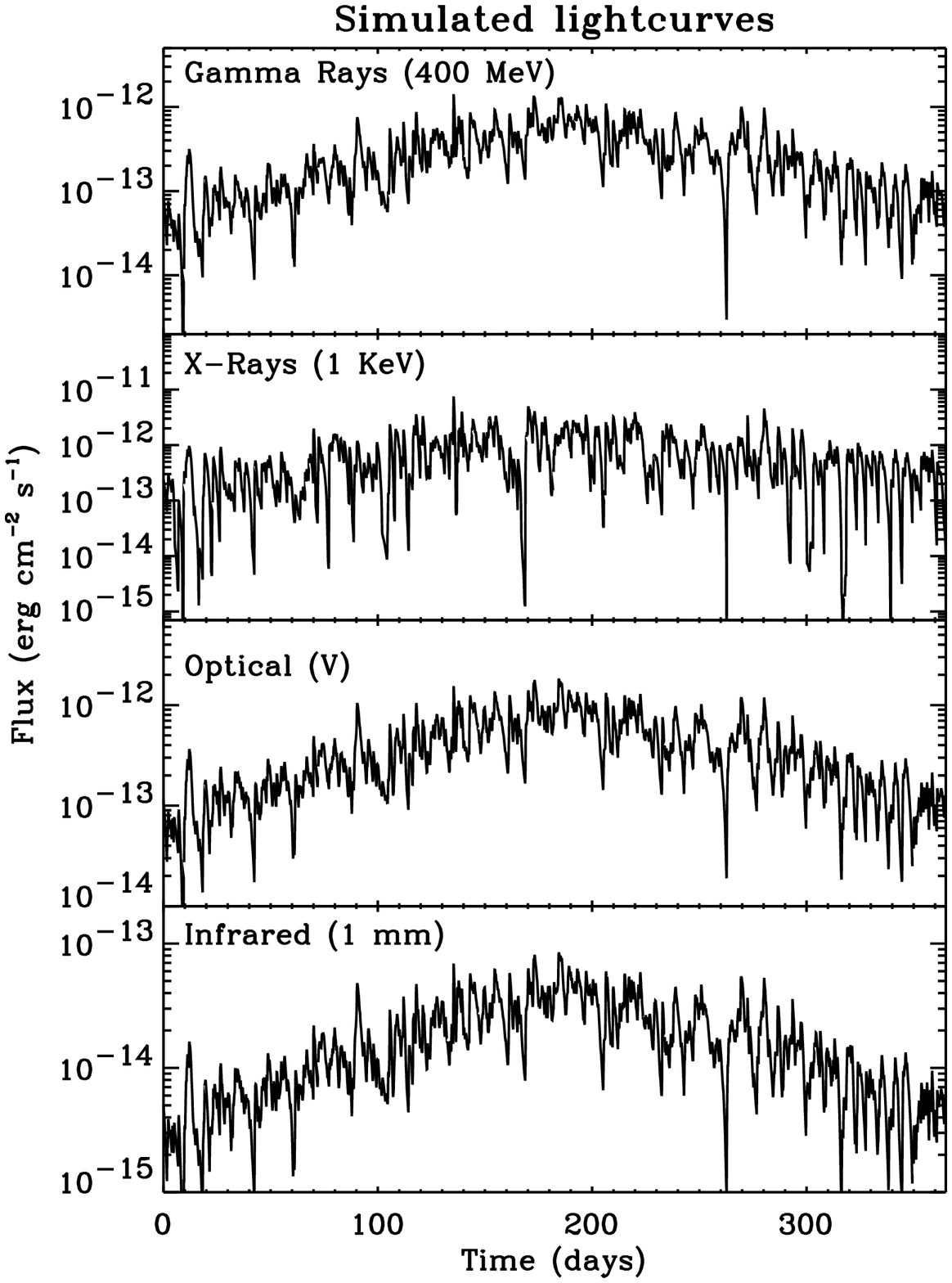,width=7cm,height=9.5true cm} } 
\vskip 0.7 true cm 
\caption{Simulated light curves for Mkn 421 in different bands, as
labeled. }
%
\vskip 1 true cm 
{\hskip 1.3 true cm 
\psfig{figure=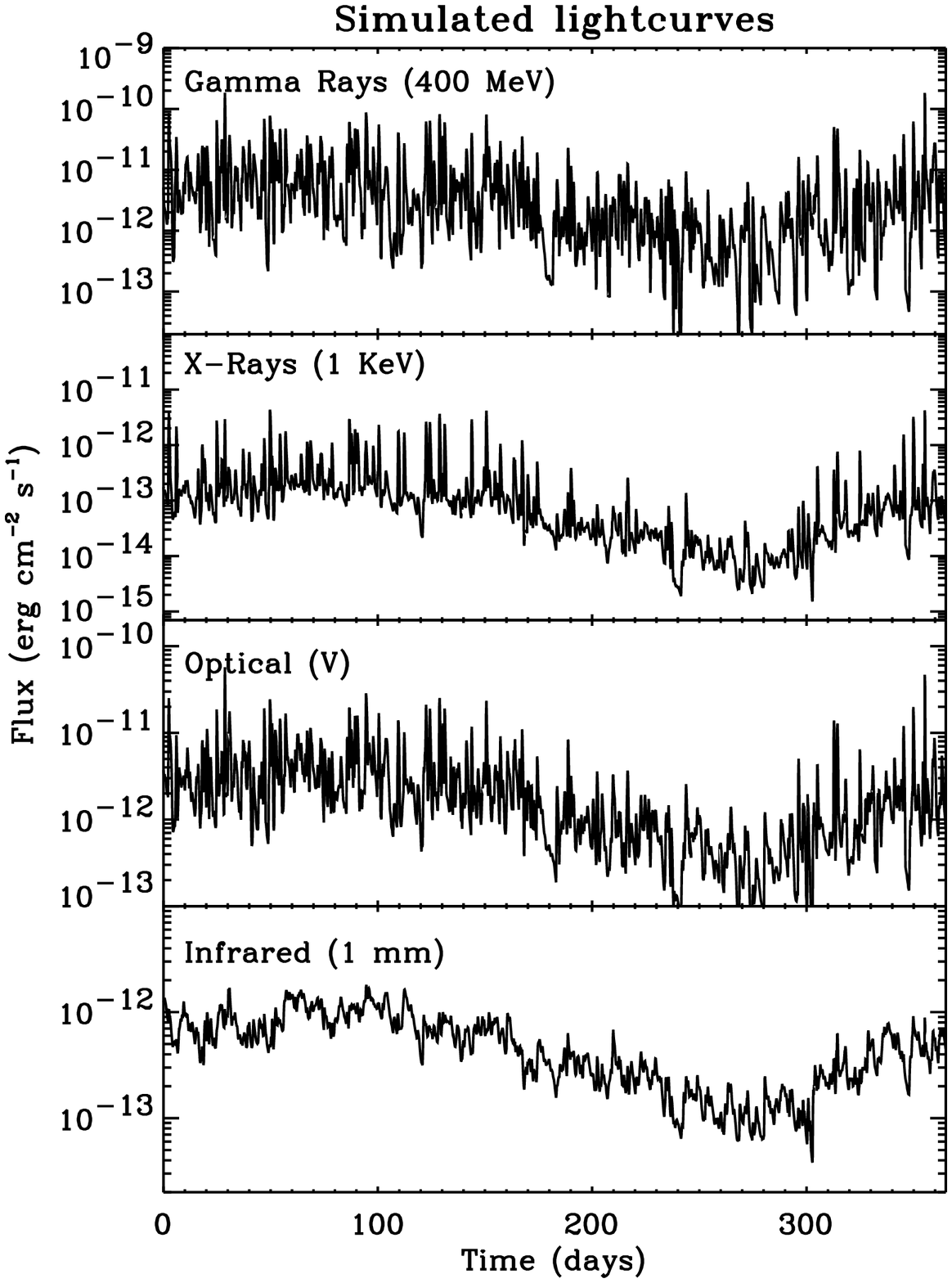,width=7cm,height=9.5true cm} } 
\vskip 0.7 true cm 
\caption{Simulated light curves for BL Lac in different bands, as
labeled. }
\end{figure} 
 
\begin{figure}[b] 
\centerline{ \psfig{figure=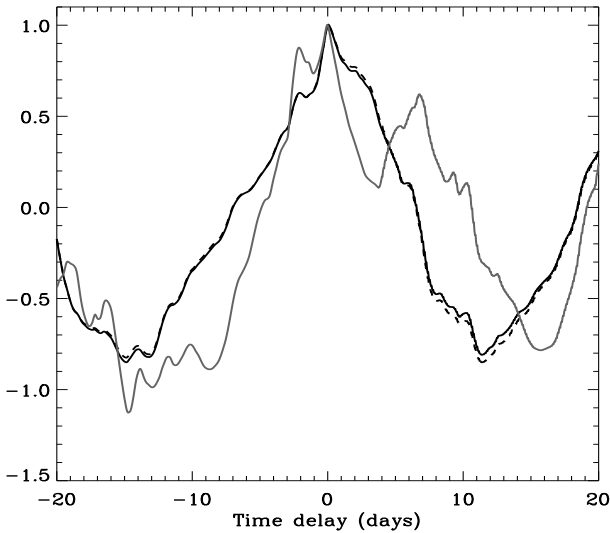,width=8cm,height=7 true cm} }
\vskip 0.3 true cm 
\caption{ Results of the cross correlations analysis for Mkn 421,
between the $\gamma$--ray and X--ray (solid grey line), optical (solid
dark line) and infrared (dashed line) light curves. There are no
delays between the different frequencies.  }
\end{figure} 
 
\begin{figure}[b] 
\centerline{ \psfig{figure=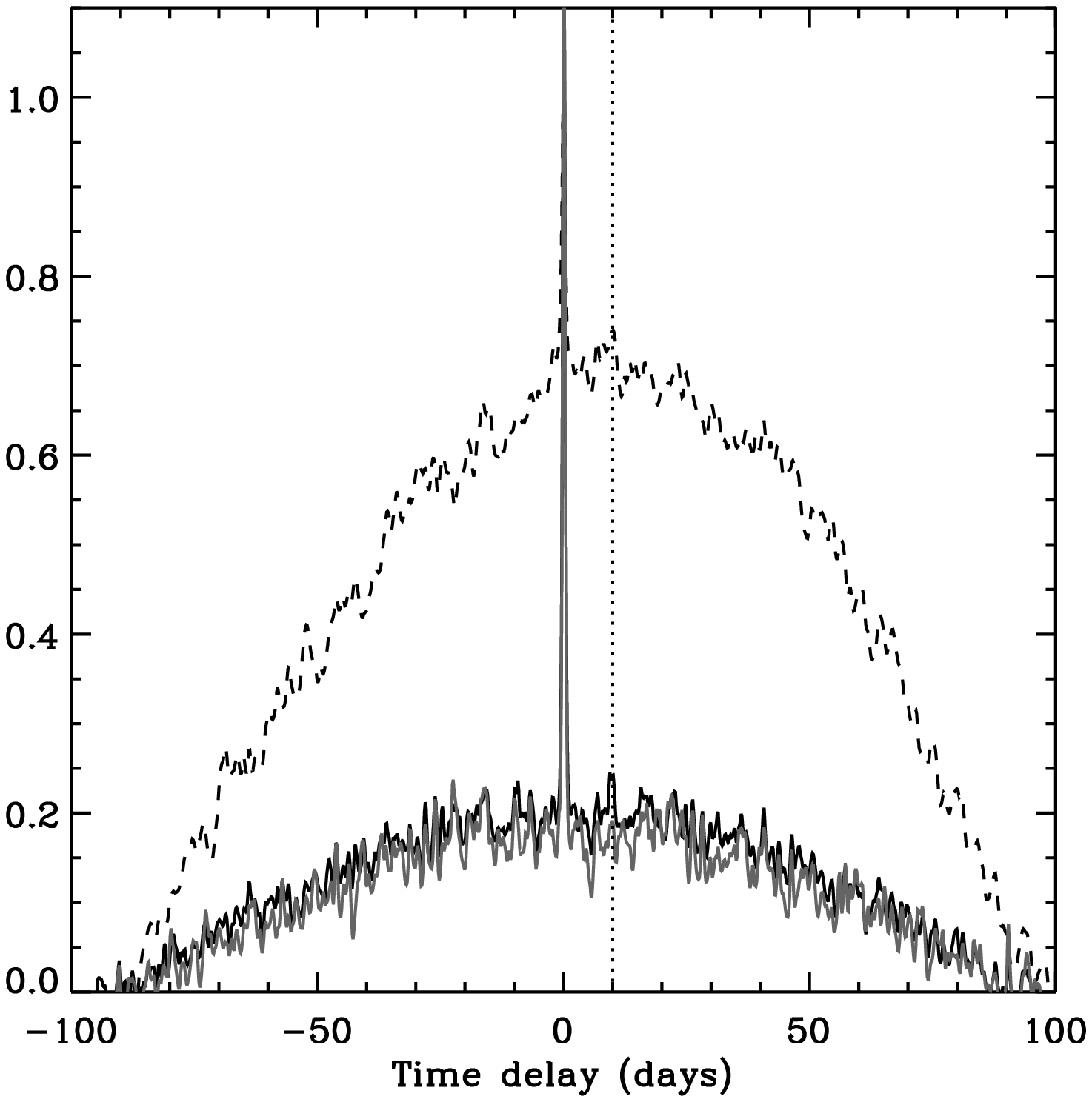,width=8cm,height=7 true cm} }
\vskip 0.3 true cm 
\caption{ Results of the cross correlations analysis for BL Lac,
between the $\gamma$--ray and X--ray (solid grey line), optical (solid
dark line) and infrared (dashed line) light curves. There is a delay
between the $\gamma$--ray and the infrared emission of roughly 10 days.
}
\end{figure} 

\vskip 0.3 true cm
\noindent {\it Light curves ---} The simulations performed aimed at
reproducing (and successfully did) all of the flux range of the SED
observed in the various multiwavelength campaigns.  However the model
also predicts the detailed variability behavior at all frequencies
(see also Tanihata et al. 2003) which has not been considered a priori
to constrain the model. Let us then examine the light curves in
different bands, reported in Fig. 9 and Fig. 10 for the two sources,
and simply compare their variability properties with observations.  A
movie showing the spectral variability during the whole simulation for
Mkn 421 (and 3C 279) can be found at the URL
http://ares.merate.mi.astro.it/$\sim$gabriele/421/index.html.  The
long term 'periodicity' that appears in the light curves is due to the
sinusoidal modulation chosen for the mass distributions of the shells.
 
In order to assess the potential of the model to reproduce the data,
we focused on comparing the simulated light curves with the observed
ones, by calculating basic indicators, namely the r.m.s., the
cumulative distribution of the variance [(Flux -- $\langle$ Flux
$\rangle)^2$], and the minimum variability doubling timescales,
defined as $\langle$ Flux $\rangle \Delta t /(\Delta$ Flux). In
particular we focused on the X--ray and optical bands for Mkn 421. For
the optical variability we sampled the simulated curves with a
frequency of about once a day for all the 300 days covered by the
simulation and compared the results with the photometric light curves
in the $V$ band presented by Tosti et al. (1998).  For the X--rays, we
considered the results presented by Tanihata et al. (2001) on 6 day
continuous observations of the source with ASCA and binned the
observed 0.2--2 keV light curve over the same time bins (0.125 days)
of the simulations.
 
In the optical band the simulated light curves show significantly
larger variability than the observed one (as from Tosti et al. 1998):
the r.m.s. from the observations is about 4 times smaller than that
derived from the model, and the Kolmogorov--Smirnoff test (KS) on the
variance distributions rejects the null hypothesis.  On the contrary,
the comparison reveals promising results for the variability in the
X--ray band.  Here, the observed and simulated r.m.s. are comparable
(0.2 vs 0.3) and the KS test shows that the variance distributions are
marginally compatible (5 per cent level).  The X--ray variability
properties of the simulated and observed light curves appear to be
similar also in terms of the variability timescale, which is of the
order of 0.3--0.4 day.
 
While a proper comparison of the variability properties would require
a more sophisticated analysis and better sampled data, these tentative
results show that the variability in the X--rays is reasonably
reproduced by the model.  A significant discrepancy appears instead in
the optical, where the model predicts larger variability: this could
be in principle improved by modifying the model parameters
(e.g. reducing the Lorentz factor contrast and modifying its
distribution), although beyond the aim of this paper. We however
notice that in the case of Mkn 421 the possible role of the host
galaxy emission in reducing the optical variability amplitude should
be assessed.
 
Finally, we computed the cross--correlations between the variability
in different energy bands. The results are presented in Fig. 11 and 12
for Mkn 421 and BL Lac, respectively.  As discussed above for Mkn 421
$\gamma_{\rm peak}$ is determined, at all radii, by $\gamma_{\rm
max}$, accounting for the absence of lags between the different
frequencies (Fig. 11).  On the contrary, in BL Lac lags ($\sim$ 10
days) are present between the infrared and the $\gamma$--ray band.  In
fact, in this source the $\gamma$--ray emission is dominated by the
external Compton within $R_{\rm BLR}$, while the infrared one by
synchrotron radiation from collisions occurring outside the broad line
region.  We remind that in the case of 3C 279 (S01) the infrared
emission is also determined by collisions outside $R_{\rm BLR}$,
resulting in lags between the IR and the $\gamma$--rays of $\sim$ 40
days.
 
\begin{table*} 
\begin{center} 
\begin{tabular}{lllllllllll} 
\hline 
Source 
& $\langle L_{\rm jet}\rangle$ 
& $\Gamma_{\rm m} $--$\Gamma_{\rm M}$ 
& $ t_{\rm v}$ 
& $ \epsilon_{\rm e}$ 
& $ \epsilon_{\rm B}$ 
& $ \zeta_{\rm e}$ 
& $ n$ 
& $ L_{\rm BLR}$ 
& $ R_{\rm BLR}$ 
& $\theta$ \\ 
&erg s$^{-1}$ & & s & & & & &erg s$^{-1}$ & cm  & degree\\ 
\hline 
Mkn 421 &8e45 &10--20 &1.3e4  &0.5 &6e-4 &8e-2  &3.1 &1e42  &1e16  &2  \\  
BL Lac  &4e46 &8--20  &1.6e4  &0.5 &1e-2 &1e-2  &4.3 &8e42  &7e16  &4  \\  
3C 279  &1e48 &10--25 &1.0e4  &0.5 &4e-3 &4e-2  &2.1 &1e45  &5e17  &2  \\ 
\hline 
\end{tabular} 
\caption{Model input parameters for Mkn 421, BL Lac and 3C279}
\end{center} 
\label{tab:fits} 
\end{table*} 
 
\section{Discussion} 
 
In this work we have considered internal shocks as the dissipation
mechanism responsible for the emission in blazars.  As mentioned in
the Introduction, this scenario is currently the most accredited to
explain the prompt gamma-ray emission in Gamma-Ray Bursts, despite of
an efficiency problem (Lazzati, Ghisellini \& Celotti 1999).  In
blazars instead the relatively low efficiency is in fact required as,
at least in powerful objects, most of the jet power is carried to the
large scales.
 
In S01 it was found that the internal shock model was successful in
reproducing the observed SED and variability properties of a powerful
blazar, namely 3C~279. However, the SED of this object does not
represent the whole of the blazar class, which covers a wide range of
spectral characteristics (luminosity, frequency of the peaks of the
emission and their relative intensity). Interestingly these parameters
appear to be correlated and the whole class can be seen as a sequence:
the frequency and the ratio of the low vs high energy peak intensity
increase with decreasing luminosity (Fossati et al. 1998).
 
We show here that the internal shock model can satisfactorily account
also for the properties of the lower power blazars. The key parameters
driving the phenomenological sequence are the jet power (proportional
to the radiated one) and the intensity of the broad lines.  These
parameters in turn regulate the SED shape, as they control the cooling
efficiency of the emitting particles. The global radiative efficiency
appears instead to be similar for all of the sources examined.
 
The internal shock scenario determines a characteristic time interval
for the injection of relativistic electrons of the order of the
dynamical timescale, when the intensity of the spectrum from each
collision is maximized.  Two regimes are relevant: {\it fast} and {\it
slow} cooling, corresponding to whether electrons of energy
$\gamma_{\rm b}$ can or cannot radiatively cool in the dynamical time.
In highly powerful blazars the fast cooling regime dominates in the
inner regions (within the BLR) where also most of the power is
dissipated.  Consequently the peak frequencies are produced by
electrons of energy $\gamma_{\rm b} m_{\rm e} c^2$.  In the weakest
blazars instead the slow cooling regime prevails over the whole
jet. Consequently only the highest energy electrons can cool over such
timescale: the peak frequencies thus shift to high values.
 
Between these two extremes there are intermediate sources, like BL
Lac, with broad lines of intermediate intensity, produced at a
distance within which a few shell-shell collisions can occasionally
take place. Observationally this corresponds to SED with moderate
Compton to synchrotron luminosity ratio, as the scattered seed photons
are only the synchrotron ones (in collisions outside the BLR).  The
rare collisions within the BLR give rise to dramatic changes in the
SED characterized by a large increase of the Compton component
(e.g. BL Lac itself, see Fig.~2).
 
The model considered here, which considers self-consistently the
dynamics and spectral emission, predicts also the time dependent
spectral properties.  In general, the selected parameters allow to
reproduce the full range of spectral 'states' observed in both BL Lac
and Mkn 421. A more detailed comparison performed for Mkn 421, and
based on the r.m.s. and variability timescales, shows good agreement
for the X--ray variability properties, while the simulated optical
variations appear to be too large with respect to the considered
observed light curve.
 
An analysis of the predicted cross-correlated variability between the
$\gamma$--ray and other bands reveals that only for BL Lac lags
($\sim$ 10 days) are expected between the $\gamma$--rays and the
infrared emission. 
 
We conclude that the internal shock scenario can account for the the
main properties of blazars and the `blazar sequence'.
 
While it has been previously pointed out that the key quantity in
reproducing the different characteristics of blazars is the ratio
between the jet and the disc (i.e. broad lines) luminosities, the
internal shock scenario discussed here provides more physical
insights, namely i) directly connects the radiated jet luminosity with
the jet effective power and ii) accounts for the `preferred' distance
where most of the luminosity is dissipated.  While the latter is
similar for high and low power blazars, the BLR is instead located at
different distances in the different sub--classes of objects (as
determined by the ionizing luminosity).
 
From a more theoretical point of view, the model does not (yet)
address the role of a seed magnetic field amplified by the shell-shell
collisions. This would probably lead to a faster synchrotron cooling
on the large scales and a flatter dependence of the $B$ field from $R$
(Fig.~3). We intend to further explore this issue.  Furthermore, only
internal shocks have been considered in this scenario. It is
conceivable that at large jet scales some entrainment may occur
causing an interaction of the jet with the external medium, possibly
leading to external shocks. As a result the radiative efficiency and
large scales synchrotron and inverse Compton emission could be
enhanced. Clearly, the inverse Compton emission could be also enhanced
if a significant photons field is present externally to the jet even
on large scales (such as microwave background and/or beamed nuclear
radiation and/or dust emission, e.g. Celotti, Ghisellini \& Chiaberge
2001; Sikora et al. 2002).
 
Observationally the improved detection sensitivity of the planned TeV
Cherenkov telescopes -- such as VERITAS, HESS, MAGIC -- will allow in
the near future to measure emission from BL Lacs in low luminosity
states, and thus to estimate their flare activity duty cycles.  Since
in the internal shock scenario the TeV emission is largely produced by
the few powerful internal collisions, it will be then possible to
further constrain model parameters such as the range of bulk Lorentz
factors and the initial separation of the shells. Analogous results
will be likely achieved for the more powerful blazars in the GeV band
thanks to the AGILE and GLAST satellites.
 
\begin{acknowledgements} 
DG thanks the Observatory of Brera in Merate for kind hospitality and
acknowledges the NSF grant AST-0307502 for financial support.  AC
thanks the JILA fellows, University of Colorado, for the warm
hospitality and the Italian MIUR and INAF for financial support.
\end{acknowledgements}

 
\end{document}